\documentclass[A4]{aa}
\def\gtsim{{_>\atop{^\sim}}}
\def\ltsim{{_<\atop{^\sim}}}
\usepackage{graphicx} 
\usepackage{amssymb}
\begin{document}
 
\title{Detection of abundant solid CO in the disk around
  \object{CRBR 2422.8-3423} \thanks{Based on observations obtained at the European Southern Observatory (ESO), Paranal, Chile, within the observing program
  164.I-0605. }}
\titlerunning{Detection of abundant solid CO toward \object{CRBR 2422.8-3423}}

\author{W.~F.~Thi\inst{1,2} \and
  K.~M.~Pontoppidan \inst{2}\and E.~F.~van~Dishoeck\inst{2}\and E. Dartois\inst{3} \and L.~d'Hendecourt\inst{3}}

\institute{Department of Physics and Astronomy, University College
  London, Gower Street, London WC1E 6BT, U.K. \and
Leiden Observatory, P. O. Box 9513, 2300 RA, Leiden, The Netherlands \and
Astrochimie Exp\'erimentale, Institut d'Astrophysique Spatiale,
  Universit\'e Paris-Sud, B\^at. 121, F-91405 Orsay, France}
\date{Received ... ; Accepted ...}
 
\abstract{We present direct evidence for CO freeze-out in a
  circumstellar disk around the edge-on class I object \object{CRBR
    2422.8-3423}, observed in the $M$ band with VLT-ISAAC at a
  resolving power $R\simeq 10\,000$.  The spectrum shows strong solid
  CO absorption, with a lower limit on the column density of 2.2
  $\times$ 10$^{18}$ cm$^{-2}$.  The solid CO column is the highest
  observed so far, including high-mass protostars and background field
  stars.  Absorption by foreground cloud material likely accounts for
  only a small fraction of the total solid CO, based on the weakness
  of solid CO absorption toward nearby sources and the absence of
  gaseous C$^{18}$O $J=2\rightarrow 1$ emission 30\arcsec south.
  Gas-phase ro-vibrational CO absorption lines are also detected with
  a mean temperature of 50 $\pm$10~K. The average gas/solid CO ratio
  is $\sim$~1 along the line of sight.  For an estimated inclination
  of 20\degr $\pm$ 5\degr, the solid CO absorption originates mostly
  in the cold, shielded outer part of the flaring disk, consistent
  with the predominance of apolar solid CO in the spectrum and the
  non-detection of solid OCN$^-$, an indicator of thermal/ultraviolet
  processing of the ice mantle. By contrast, the warm gaseous CO
  likely originates closer to the star.  }\maketitle

\section{Introduction}

Interstellar gas and dust form the basic ingredients from which
planetary systems are built (e.g., van Dishoeck \& Blake \cite{vDB98},
Ehrenfreund \& Charnley \cite{Ehrenfreund00}). In particular, the icy
grains can agglomerate in the cold midplane of circumstellar disks to
form planetesimals such as comets. In the cold ($T<$~20~K) and dense
($n_{\rm H} = 10^6-10^9$ cm$^{-3}$) regions of disks, all chemical
models predict a strong freeze-out of molecules onto grain surfaces
(e.g., Aikawa et al.\ \cite{Aikawa02}).  The low molecular abundances
in disks compared to those in dense clouds as derived from
(sub)millimeter lines is widely considered to be indirect evidence for
freeze-out (Dutrey, Guilloteau \& Gu\'elin\ \cite{Dutrey97}; Thi et
al.\ \cite{Thi01}).
 
Observations of gaseous and solid CO have been performed for a few
transitional objects from class I to class II that are known to posses
a disk. Boogert et al.\ (\cite{Boogert02a}) observed \object{L~1489}
in Taurus -- a large 2000 AU rotating disk --, but the amount of solid
CO is not exceptionally high ($\sim$7\% of gaseous CO). This may stem
from the fact that these systems are still far from edge-on
(inclination $\sim 20{\degr}-30{\degr}$) so that the line of sight
does not intersect the midplane, the largest reservoir of solid CO.
Shuping et al.\ (\cite{Shuping01}) found strong CO depletion toward
\object{Elias 18} in Taurus, but both the disk stucture and its
viewing angle are not well constrained.  More promising targets are
pre-main-sequence stars for which near-infrared images have revealed
nebulosities separated by a dark lane (e.g. Padgett et al.\ 
\cite{Padgett99}).  The lane is interpreted as the cold midplane of a
disk seen close to edge-on where visible and even near-infrared light
are extremely extinct. Among such dark-lane objects, \object{CRBR
  2422.8-3423} is a red ($H-K$=4.7) low luminosity ($L=0.36~L_{\sun}$,
Bontemps et al.\ \cite{Bontemps01}) object surrounded by a near
edge-on disk, discovered in images with the ESO {\it Very Large
  Telescope} (VLT) at 2 $\mu$m (Brandner et al.\ \cite{Brandner00}).
Its spectral energy distribution (SED) is consistent with that of a
class I object or an edge-on class II object with strong silicate
absorption at 9.6~$\mu$m. It is located at the edge of the
\object{$\rho$ Oph cloud} complex in the core F, $\sim$ 30\arcsec west
of the infrared source \object{IRS~43} and a few arcmin south-east of
\object{Elias 29} (Motte, Andr\'e \& Montmerle \cite{Motte98}).

This letter reports the detection of a large quantity of solid CO and
the presence of gaseous CO in the line of sight of \object{CRBR 2422.8-3423}
using the ESO-VLT (\S 2 and 3). Possible contamination by foreground
cloud material is considered in \S~\ref{discussion}, followed by a
discussion on the location and origin of the CO gas and dust in the
disk (\S 5).
   
\section{Observations}\label{observations}
 
\object{CRBR 2422.8-3423} was observed with the ESO VLT-ANTU mounted
with the {\it Infrared Spectrometer And Array Camera} (ISAAC) on May
6, 2002.  A spectrum at $R\approx 10^4$ was obtained in the $M$ band
using a slit-width of 0\farcs 3, which matched the excellent seeing
($\simeq$~0\farcs 3 at 4 $\mu$m).  The integration time was 36
minutes, resulting in a continuum $S/N$ of $\simeq$~20.  The spectra
of \object{BS~6084} (B1III) and \object{BS~6378} (A2.5Va), observed
immediately before and after \object{CRBR 2422.8-3423}, were used to
remove the atmospheric features and to calibrate the spectrum in flux
and wavelength. All the reduction steps were carried out using an
in-house data reduction package written in {\it IDL}.

Observations of gaseous CO (sub)millimeter lines on source and at an
offset position 30\arcsec south were obtained at the {\it James Clerk
  Maxwell Telescope} (JCMT)\footnote{\footnotesize The James Clerk
  Maxwell Telescope is operated by the Joint Astronomy Centre in Hilo,
  Hawaii on behalf of the Particle Physics and Astronomy Research
  Council in the United Kingdom, the National Research Council of
  Canada and The Netherlands Organization for Scientific Research.} in
June 2002.  The dual-polarization receiver B3 was tuned to observe the
$^{12}$CO $J=3\rightarrow 2$ (345.796~GHz, beam 13\farcs 8) and
$^{13}$CO $J=3\rightarrow 2$ (330.587~GHz, beam 14\farcs 4) lines, and
receiver A3 to obtain C$^{18}$O $J=2\rightarrow 1$ (219.560 GHz, beam
22\arcsec.2).  The backend was the Digital Autocorrelator
Spectrometer, set at a resolution of $\sim$0.15 km s$^{-1}$.  The beam
efficiencies are $\eta_{\rm mb}$=0.63 (345 GHz) and $\eta_{\rm
  mb}$=0.69 (230 GHz).  The fluxes were calibrated against the bright
nearby source \object{IRAS 16293-2422}. The observations were acquired
in position switching with a throw of 1800\arcsec south and reduced
using the SPECX software.

\section{Results and analysis}\label{results}

\subsection{VLT-ISAAC spectrum}

\begin{figure}
  \resizebox{\hsize}{!}{\includegraphics[]{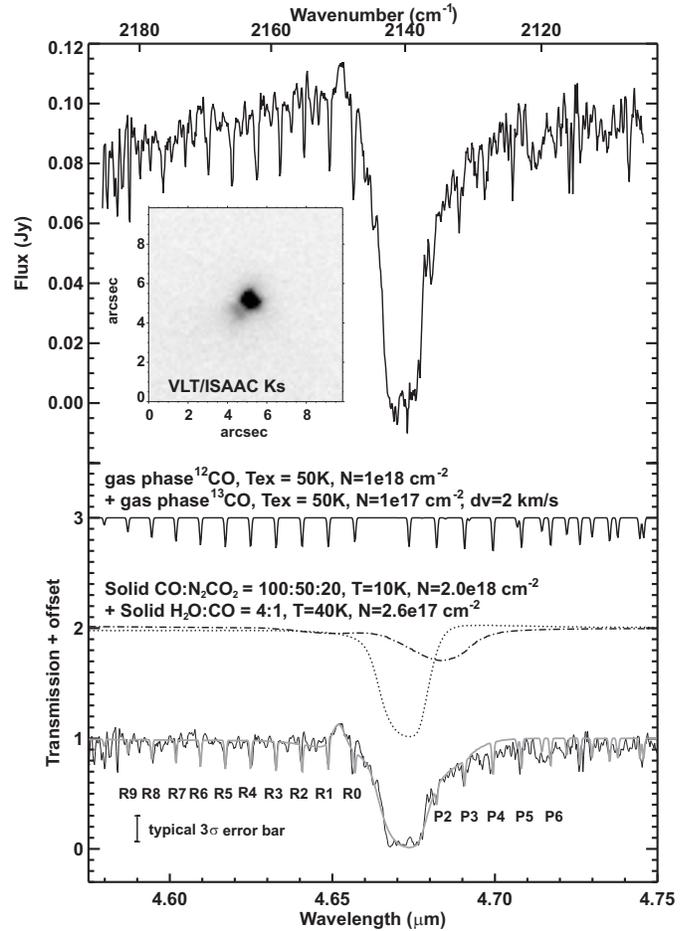}}
\caption{VLT-ISAAC $M$band spectrum toward \object{CRBR 2422.8-3423} (upper
  panel). The insert shows the VLT Ks image taken by Brandner et al.\ 
  (\cite {Brandner00}). The upper curve of the lower panel shows the
  gas-phase model, convolved to the observed spectral resolution and
  shifted in wavelength to account for a source heliocentric velocity
  of 10 km s$^{-1}$.  The middle curve displays laboratorium
  absorption spectra of apolar (dotted) and polar (dashed) CO ice
  mixtures. The sum of the two models including Pfund $\beta$ emission
  at 4.65 $\mu$m is overlaid on the observed spectrum in the bottom
  curve.}
\label{fig_crbr_vlt} 
\end{figure}

A saturated absorption feature centered at $\sim$ 4.67~$\mu$m and
assigned to solid CO is seen in the $M$-band spectrum displayed in the
upper panel of Fig.\ \ref{fig_crbr_vlt}. No broad absorption at 4.62
$\mu$m, usually attributed to solid OCN$^-$ and a signpost of thermal
and/or ultraviolet processing, is present, although the limit of $\tau
\leq$~0.05 is not strong. Unresolved gas-phase CO lines are also
detected. A continuum which takes into account absorption by silicate
is fitted to the data. The continuum substracted spectrum is shown in
the lower panel of Fig.\ \ref{fig_crbr_vlt} in transmission scale.

The best fit to the saturated solid CO band is $\tau \approx 6.2$,
corresponding to $N$(CO$_{\rm ice})$ $\approx 2.2 \times 10^{18}$
cm$^{-2}$ using the integrated band strength of Gerakines et al.\ 
(\cite{Gerakines95}).  This is the highest solid CO column density
found to date, even compared with ice-rich high-mass sources such as
\object{NGC 7538 IRS9} (e.g.\ Chiar et al.~\cite{Chiar98},
\cite{Chiar95}). The solid CO profile toward \object{CRBR 2422.8-3423}
can be decomposed into a narrow saturated component, usually ascribed
to apolar CO in a H$_2$O-poor matrix, and a broad red wing at 4.685
$\mu$m, indicative of polar CO in an H$_2$O-rich matrix.  Using
optical constants from the Leiden database including grain shape
effects (Ehrenfreund et al.\ \cite{Ehrenfreund97}), a $\chi ^2$ search
for the best-fitting laboratory mixtures was performed.  The
saturation of the line prevents any unique fit, but the apolar CO is
best matched by a mixture CO:N$_2$:CO$_2$=100:50:20 at 10~K with
$N$=2.0 $\times$ 10$^{18}$ cm$^{-2}$. The red-wing is fitted by
H$_2$O:CO = 4:1 at 40~K with 2.6 $\times$ 10$^{17}$ cm$^{-2}$.  The
latter column density is comparable to that found toward L~1489, which
has a luminosity of 3.7 $L_{\odot}$.  The large amount of apolar CO
indicates low temperatures along the line of sight, since it
evaporates around $\sim 20$~K.

The VLT spectrum shows the presence of narrow gas-phase $^{12}$CO and
$^{13}$CO ro-vibrational lines originating from levels up to $J=9$
(250 K above ground).  Synthetic LTE model spectra were fitted to the
observed spectrum using data from the HITRAN database (Rothman et al.\ 
\cite{Rothman97}).  The fit parameters are the gas temperature $T_{\rm
  ex}$, the column density $N_{\rm gas}$(CO) and the velocity
broadening $\Delta V$, assumed to be smaller than 2~km~s$^{-1}$ from
the JCMT data.  The limited number and the high optical depth of the
$^{12}$CO lines prevent a unique fit, but the best result (see Fig.~1)
is obtained for $T_{\rm ex}=40-60$~K, which is probably a mean between
the cold and warm components along the line of sight. The fit includes
the $^{13}$CO lines.  The optical depth of the $^{12}$CO lines is
likely underestimated so that the gas-phase CO column density can
range from 1 to 6 $\times$ 10$^{18}$ cm$^{-2}$.  Adopting a mean value
of 3 $\times$ 10$^{18}$ cm$^{-2}$, the line of sight average gas/solid
CO ratio is $\sim$~1. Comparing the column of gas-phase CO with that
of polar solid CO only, the ratio drops to $\sim$~0.1. The latter value
is comparable to that found for \object{L~1489} (0.07, see Boogert et
al.\ \cite{Boogert02a}) but still higher than for \object{Elias 18}
(0.01, Shuping et al.\ \cite{Shuping01}). If the lines were
significantly wider than 2 km s$^{-1}$ as found for \object{L~1489}
($\sim$~20~km~s$^{-1}$), the gaseous CO column density would drop by
an order of magnitude.

\subsection{JCMT spectra}

\begin{figure}
\resizebox{\hsize}{!}{\includegraphics[angle=90]{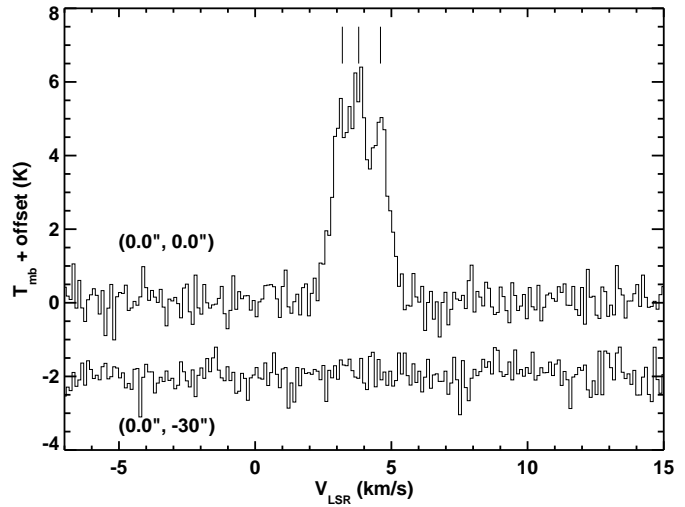}}
\caption{C$^{18}$O $J=2\rightarrow 1$ obtained using the JCMT toward
  \object{CRBR 2422.8-3423} on source (upper spectrum) and at an
  offset position 30$''$ south (lower spectrum). }
\label{fig_crbr_jcmt}
\end{figure}

The C$^{18}$O $J=2\rightarrow 1$ emission observed with the JCMT is
shown in the upper panel of Fig.\ \ref{fig_crbr_jcmt}. Strong
$^{12}$CO $J=3\rightarrow 2$ ($\sim 12$~K peak temperature) and
$^{13}$CO $J=3\rightarrow 2$ ($\sim 7$~K peak temperature) lines are
also detected, but are not displayed here since their profiles are
similar to that of C$^{18}$O.  C$^{18}$O $J=2\rightarrow 1$ is
detected on source only.  Compared with single-dish CO 3--2 and 2--1
lines from disks around older isolated pre-main-sequence stars in
Taurus (e.g., Thi et al.\ \cite{Thi01}), the lines toward \object{CRBR
  2422.8-3423} are more than an order of magnitude stronger and do not
show the double-peak profile resulting from the projection of a disk
in Keplerian rotation.  The \object{CRBR 2422.8-3423} spectrum shows
three peaks, two of which have velocities similar to those seen toward
\object{Elias 29} and interpreted as arising from the cloud ridge in
which the source is embedded ($N({\rm CO_{gas}})_{\rm
  cloud}\simeq$~2.9~$\times$~10$^{18}$ cm$^{-2}$, Boogert et al.\ 
\cite{Boogert02b}). The total integrated C$^{18}$O 2--1 intensity on
source is 13.6 $\pm$ 4 K km s$^{-1}$. The corresponding $^{12}$CO
column density is (0.5--1) $\times$ 10$^{19}$ cm$^{-2}$ assuming
$T_{\rm ex}=15$~K (cf.\ Motte et al.\ \cite{Motte98} for $\rho$ Oph
clump F) and $^{18}$O/$^{16}$O=560 (Wilson \& Rood \cite{Wilson94}).
For $T_{\rm ex}=50$~K, the column densities are lowered by 30\%.  The
non-detection of C$^{18}$O $J=2\rightarrow 1$ at the offset position
suggests a $^{12}$CO column density less than $1 \times 10^{17}$
cm$^{-2}$ for $T_{\rm ex}=15$~K , around 20 times lower.

\section{Contamination by foreground material?}\label{discussion}

Absorption studies toward young stellar objects located in the
\object{$\rho$ Oph cloud} complex can be dominated by foreground
cloud(s) (Boogert et al.\ \cite{Boogert02b}).  The SCUBA 850$\mu$m map
of $\rho$ Oph by Johnstone et al.\ (\cite{Johnstone00}) shows that
\object{CRBR 2422.8-3423} is indeed at the edge of the ridge which
contains both Elias 29 and the nearby class I object \object{IRS~43}.
The lack of CO emission at the offset position indicates however that
the bulk of the gas-phase CO is located within 3000 AU from
\object{CRBR 2422.8-3423}.  The similarity of the gaseous CO column
densities derived from the on-source C$^{18}$O 2--1 emission and the
VLT-ISAAC infrared absorption may be fortuitous since the VLT-ISAAC
data do not probe very cold CO with small ($\ltsim 1$ km s$^{-1}$)
line widths. We cannot exclude that part of the gas-phase CO seen in
the infrared arises in a more extended envelope or cloud, but the fact
that the CO excitation temperature is significantly above 10~K
indicates that at least some fraction must originate close to the
young star in a warm part of the disk.

For solid CO, there are strong arguments that most of the absorption
must arise in the disk. The bright nearby ($\sim$~30\arcsec east)
source \object{IRS~43} was observed simultaneously with \object{CRBR
  2422.8-3423} with VLT-ISAAC and has a CO ice optical depth of only
1.87 $\pm$ 0.02, corresponding to $N$(CO$_{\rm
  ice}$)=7~$\times$~10$^{17}$ cm$^{-2}$ (Pontoppidan et al., in
prep.). This is at least a factor of 3 lower than toward \object{CRBR
  2422.8-3423} even though its 850 $\mu$m flux is a factor of 1.5
higher.  Toward \object{Elias~29}, solid CO has an optical depth of
only 0.25, most of which is believed to be located in the foreground
clouds.  Compared to the optical depth found toward \object{CRBR
  2422.8-3423} ($\tau_{\rm ice}$(CO)~$\gtsim$~6), this amount of
foreground material can account for only an insignificant fraction of
the observed solid CO.  Finally, if clouds happen to lie in front of
\object{CRBR 2422.8-3423}, the moderate extinction ($A_{\mathrm
  V}<$10) of those clouds probably prevents them to harbour
significant amounts of solid CO (Shuping et al.\ \cite{Shuping00}).
Indeed, in a $M$-band survey of more than 30 young stellar objects,
this is the deepest CO ice band observed.

\section{Discussion}

\begin{figure}
\resizebox{\hsize}{!}{\includegraphics[]{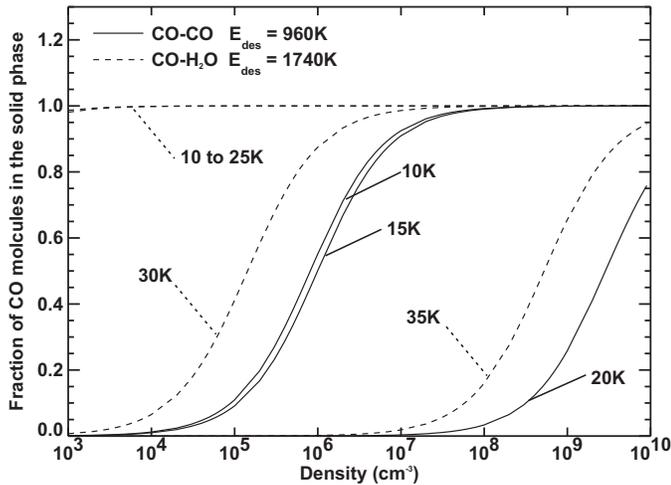}}
\caption{Fraction of CO molecules in the solid phase at
  different temperatures as function of density. 
  Adsorption onto H$_2$O and CO ice
  produce different results. The desorption energies are from Sandford
  \& Allamandola (\cite{Sandford88a}) for CO-H$_2$O and
  Sandford et al.\  (\cite{Sandford88b}) for CO-CO.}
\label{fig_co_fraction}
\end{figure}

A simple disk model cf.\ Chiang \& Goldreich (\cite{CG99}) with
$T_*$=3500~K and a disk radius of 250~AU has been adopted to
investigate the location of the solid and gaseous CO seen in infrared
absorption.  Because of its higher temperature, the gaseous CO is
likely not co-located with the bulk of the apolar solid CO, which
evaporates at $\sim$~20~K. The polar solid CO can however reside in
the same region of the disk as the CO gas at 40--60~K.  Assuming that
CO is frozen out at $<$~20~K (apolar CO) and $<$~40~K (polar CO), the
best fit to the column densities, gas/solid CO and polar ice/apolar
ice ratio is obtained for $i$=20$\pm 5\degr$.  This inclination is
consistent with the flux asymmetry seen in the near-infrared image of
Brandner et al.\ (\cite{Brandner00}). For such line of sight, most of
the CO ice is located above the midplane in the outer disk, whereas
the CO gas is found in the warm inner disk.  Thus, the overall CO
depletion could be significantly higher than the ratio of $\sim$~1
found here.
  
Several time-dependent chemical models were run to quantify the
gas/solid CO ratios in different density and temperature regimes.  The
models simulate gas-phase chemistry, freeze-out onto grain surfaces
and thermal as well as non-thermal evaporation.  Cosmic-ray induced
desorption is modeled as in Hasegawa \& Herbst (\cite{Hasegawa93}),
which may be an overestimate for large grains in disks (Shen et al.,
in prep.).  The sticking coefficient was set at 0.3 to account for
other non-thermal mechanisms (e.g., Schutte \& Greenberg
\cite{Schutte97}) and photodesorption is assumed ineffective.  At
$T>T_{\rm evap}$ thermal desorption dominates, whereas at $T<T_{\rm
  evap}$ cosmic-ray desorption prevails.  In the model, $T_{\rm evap}$
is 20~K for apolar CO ice ($E_{\rm des}$=960~K) and 40~K ($E_{\rm
  des}$=1740~K) for polar CO. To illustrate the effects of thermal
and cosmic-ray induced desorptions, Fig.\ \ref{fig_co_fraction} shows
the gas/solid CO ratio at chemical equilibrium. In cold ($T<$~20~K)
but moderately dense (10$^4-10^5$ cm$^{-3}$) regions of the disk
around \object{CRBR 2422.8-3423}, only a small amount of CO is
depleted onto grains in the apolar form, but much larger fractions
$>$~50\% can occur at higher densities ($\sim 10^6-10^8$ cm$^{-3}$).

In summary, we detected a large amount of solid CO in the line of
sight toward \object{CRBR 2422.8-3423}. The majority of this ice is
likely located in the flaring outer regions of the edge-on
circumstellar disk.  Very high resolution ($R>10^5$) near-infrared
spectroscopy is needed to reveal the gaseous CO line profiles and thus
their origin and the gas dynamics in the inner disk.  Future
submillimeter interferometer data can probe the velocity pattern and
excitation conditions of the gas as functions of disk radius, whereas
mid-infrared spectroscopy with, e.g., the {\it Space Infrared
  Telescope Facility} (SIRTF) will allow searches for other ice
components, in particular solid CO$_2$.

\begin{acknowledgements}
  WFT thanks PPARC for a postdoctoral grant to UCL. Astrochemistry in
  Leiden is supported by a Spinoza grant from the Netherlands
  Organization for Scientific Research (NWO) and a PhD grant from the
  Netherlands Research School for Astronomy (NOVA). We thank the ESO
  staff for their help during the observations, P.~Papadopoulos for
  performing the JCMT observations and D.~Johnstone for a blow-up of
  the SCUBA map of $\rho$ Oph.
\end{acknowledgements}

\end{document}